\documentclass[namedreferences,natbib]{solarphysics}
\usepackage[optionalrh,solaromanenum]{spr-sola-addons} 
\usepackage{graphicx}                    
\usepackage{amssymb}                    
\usepackage{color}                       
\usepackage{url}                         
\usepackage[pdfborder={0 0 0},colorlinks=true,linkcolor=black,citecolor=black,filecolor=black,urlcolor=blue,breaklinks=true,dvipdfm]{hyperref} 

\usepackage{solaheader}	
\newcommand{\solareceiveddate}{2 October 2012} 
\newcommand{\solaaccepteddate}{2 March 2013}
\begin{document}

\begin{article}

\begin{opening}

\title{Interplanetary Nanodust Detection by the \textit{Solar Terrestrial Relations Observatory}/WAVES Low Frequency Receiver}

%
\author{G.~\surname{Le Chat}$^{1,2,3}$\sep
        A.~\surname{Zaslavsky}$^{3}$\sep
        N.~\surname{Meyer-Vernet}$^{3}$\sep
        K.~\surname{Issautier}$^{3}$\sep
        S.~\surname{Belheouane}$^{3}$\sep
        F.~\surname{Pantellini}$^{3}$\sep
        M.~\surname{Maksimovic}$^{3}$\sep
        I.~\surname{Zouganelis}$^{4}$\sep
        S.D.~\surname{Bale}$^{5}$\sep
        J.C.~\surname{Kasper}$^{1,2}$      
       }

%
\runningauthor{G. Le Chat {\it et al.}}
\runningtitle{Interplanetary Nanodust Detection by STEREO/WAVES LFR}

%
  \institute{$^{1}$ Harvard-Smithsonian Center for Astrophysics, Cambridge, USA\\
                     email: \url{glechat@head.cfa.harvard.edu}\\ 
             $^{2}$ NASA Lunar Science Institute, Moffet Field, CA, USA\\
             $^{3}$ LESIA, Observatoire de Paris, CNRS, UPMC, Universit\'e
Paris Diderot; 5 Place Jules Janssen, 92195 Meudon, France\\
             $^{4}$ LPP, UPMC, \'Ecole Polytechnique, CNRS, 4 av. de Neptune, 94107 Saint-Maur-des-Foss\'es, France\\
             $^{5}$ Space Science Laboratory, University of California, Berkeley, USA
}

\begin{abstract}
New measurements using radio and plasma-wave instruments in interplanetary space have shown that nanometer-scale dust, or nanodust, is a significant contributor to the total mass in interplanetary space.  Better measurements of nanodust will allow us to determine where it comes from and the extent to which it interacts with the solar wind. When one of these nanodust grains impacts a spacecraft, it creates an expanding plasma cloud, which perturbs the photoelectron currents. This leads to a voltage pulse between the spacecraft body and the antenna. Nanodust has a high charge/mass ratio, and therefore can be accelerated by the interplanetary magnetic field to speeds up to the speed of the solar wind: significantly faster than the Keplerian orbital speeds of heavier dust.  The amplitude of the signal induced by a dust grain grows much more strongly with speed than with mass of the dust particle.  As a result, nanodust can produce a strong signal, despite their low mass. The WAVES instruments on the twin {\it Solar TErrestrial RElations Observatory} spacecraft have observed interplanetary nanodust particles since shortly after their launch in 2006. After describing a new and improved analysis of the last five years of STEREO/WAVES \textit{Low Frequency Receiver} data, a statistical survey of the nanodust characteristics, namely the rise time of the pulse voltage and the flux of nanodust, is presented. Agreement with previous measurements and interplanetary dust models is shown. The temporal variations of the nanodust flux are also discussed.
\end{abstract}

%
\keywords{Interplanetary Dust, Nanodust, In situ Dust Detection, Radio Antennas, STEREO/WAVES}

\end{opening}

%
 \section{Introduction}
 
 Radio and plasma-wave instruments have measured interplanetary dust {\it in situ} since 1983. Dust grains were first detected with electric sensors in planetary environments, during the G-ring crossing of Saturn by {\it Voyager 2} \citep{Aubier83, Gurnett83}. Dusts have been measured {\it in situ} by such instruments in planetary environments, during the ring--plane crossings of Uranus \citep{Meyer86, Gurnett87} and Neptune \citep{Pedersen91, Gurnett91}, the E-ring crossing of Saturn by {\it Voyager 1} \citep{Meyer96}, and recently by \textit{Cassini} \citep{Kurth06}. Dusts have also been observed in cometary environments, with the {\it International Cometary Explorer} spacecraft, which crossed the tail of comet Giacobini--Zinner in 1985 \citep{Gurnett86}. Later, the {\it Vega} spacecraft measured dust particles near comet Halley \citep{Oberc90, Oberc92, Oberc93}. 
 
In addition to the measurements near outer planets by classical cosmic dust analyzers  (see \citet{Hsu12} and references therein), nanosized dust grains have been measured recently in the vicinity of Jupiter by {\it Cassini} \citep{Meyer09b}, and in the solar wind at 1 Astronomical Unit (AU) \citep{Meyer09,Zaslavsky12, Meyer12}.
 
 In this article, we present an improved analysis of the detection of nanometer dust grains by the  {\it Solar TErrestrial RElations Observatory}/WAVES \textit{Low Frequency Receiver} (LFR) (Section \ref{s:method}), from whose measurements we deduce nanodust characteristics (Section \ref{s:dust}). We study five years of data between 2007 and 2011. This data set corresponds to $3\,980\,919$ spectra for  STEREO-Ahead (A) and $3\,952\,952$ for STEREO-Behind (B). This extensive study is obtained in the light of the recently proposed scenario for the interpretation of the nanodust-impact-associated potential pulses measured between the spacecraft main body and the boom antennas \citep{Zaslavsky12, Pantellini12b, Pantellini12}. This scenario was enabled by the detailed analysis of the \textit{Time Domain Sampler} (TDS: \citealt{Zaslavsky12}) data. 

 \section{Dust Impact Detection by STEREO/WAVES LFR}\label{s:method}
 
 \subsection{Description of the Instrument}
 
 The STEREO-A and STEREO-B spacecraft travel close to the Earth's orbit, STEREO-A preceding Earth and STEREO-B trailing behind  at angular distances increasing by $42$ degrees per year. The STEREO/WAVES or S/WAVES instrument \citep{Bougeret08} is the radio and plasma wave experiment aboard the STEREO spacecraft. Its sensors consist of three, mutually orthogonal, six-meter long monopole antennas, which we label X, Y, and Z  \citep{Bale08}. The S/WAVES instrument produces power spectra of electric field fluctuations (\textit{Low Frequency Receiver}: LFR, and \textit{High Frequency Receiver}: HFR), along with limited samples of the raw voltage timeseries, also called waveforms (\textit{Time Domain Sampler}: TDS). In this study, we will focus on the LFR data measured when high-velocity nanodust particles impact the spacecraft. An extensive study of the TDS data during dust events was made by \citet{Zaslavsky12}. In dipole mode, LFR measures the difference between the voltages of the X and Y antennas about every 40 seconds, which is the typical time resolution for both STEREO spacecraft. The LFR frequency range is divided in three frequency bands of two octaves: A, B, and C. For each band, the acquisition time of the power spectrum is inversely proportional to the frequency at the center of the band. Therefore, band A (2.5 kHz \,--\, 10 kHz) has the greatest acquisition time (406 ms, respectively 4 times and 16 times greater than for bands B and C), which makes it the most sensitive to dust impacts, due to their non-stationary behavior and to the automatic gain control.
 
 \subsection{Analysis of the Dust Impact Signal}
 
For dust grains impacting the spacecraft, with rate $N$, voltage pulse maximum amplitude $\delta V$, and rise time $\tau$ \citep[Figure 2]{Zaslavsky12}, the theoretical power spectrum is \citep{Meyer85}
 \begin{equation}\label{v2f}
  V_\mathrm{f}^2\ \approx\ 2\langle N\delta V^2 \omega^{-2}(1\,+\,\omega^2 \tau^2)^{-1}\rangle
 \end{equation}
 \noindent at frequencies [$f\,=\,\omega/2\pi$] much greater than the pulses' inverse decay time (itself much greater than $\tau$), where the angular brackets stand for averaging over the pulses detected during the acquisition time. This produces a power spectrum varying as $f^{-4}$ at frequencies greater than $1/2\pi \tau$, and as $f^{-2}$ at frequencies lower than $1/2\pi \tau$. Comparing Equation (\ref{v2f}) to the LFR band A measurements, it is possible to determine the rise time [$\tau$] and the amplitude [$\langle N\delta V^2 \rangle$]. To do so, it is necessary to evaluate the ubiquitous thermal noise \citep{Meyer89}, which decreases less steeply than the spectrum produced by dust impacts. The relative intensity of signals from dust and thermal noise is such a strong function of frequency that band C, which responds to the highest-frequency signals, only detects thermal noise, even during dust impacts \citep{Meyer09}. It is noteworthy that contrary to TDS, LFR cannot measure individual impacts \citep{Meyer09}, due to: i) the Automatic Gain Control (AGC), which determines the receiver gain \citep{Bougeret08} and which needs multiple dust impacts to record them, ii) the fact that LFR integrates all of the impacts occurring during its acquisition time, and iii) the wavelet-like transform \citep{Sitruk95}, which is used on board to compute the signal power spectral densities. 

 
 
 \begin{figure}
	\centerline{ \includegraphics[width=0.8\textwidth]{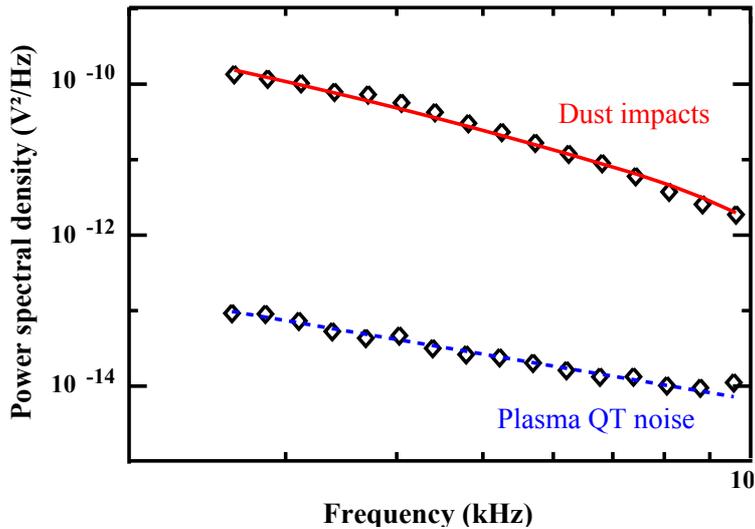}}
 	\caption{Power spectrum measured on Band A by STEREO-A on 10 October 2009 at 5:12 (upper spectrum) and 7:25 (lower spectrum). The upper spectrum is typical of dust measurement, whereas the lower one corresponds to the plasma quasi-thermal noise measurement. The lines correspond to the best fit, in case of dust (red solid line) and plasma thermal noise (blue dashed line).}\label{f:fit}
 \end{figure}

Figure \ref{f:fit} shows the power spectrum measured in Band A by STEREO-A and the result of our fitting procedure in the case of dust impacts (upper spectrum) and when measuring the ubiquitous plasma quasi-thermal noise (lower spectrum). This figure illustrates the difference in power level and spectral index between the plasma thermal noise and the detection of nanodusts at these frequencies. In the case of dust impacts, we fit Equation (\ref{v2f}) to the power spectrum measured in Band A to obtain the two free parameters: the rise time $\tau$ and the amplitude $\langle N\delta V^2 \rangle$. 

 Previous studies have shown that on STEREO-A the X-antenna is in the best geometric configuration to directly detect the plasma cloud from fast nanometer-scale dust impacts through the perturbation of the photoelectron population around the antenna \citep{Pantellini12, Zaslavsky12}. Consequently, the voltage on the X-antenna is typically 20 times larger than on the Y antenna \citep{Zaslavsky12}. The power spectrum measured by LFR in Band A (based on the difference in voltage between the X- and Y-antennas) is then much higher than the thermal noise, at least two orders of magnitude higher. The voltage pulse on the X-antenna is given by \citep{Zaslavsky12}
 \begin{equation}\label{deltaV}
	\delta V_\mathrm{X}\,\approx\,\frac{\Gamma T l}{L}
 \end{equation}
 
 \noindent where $\Gamma\,\approx\,0.5$ and $L\,=\, 6\,\mathrm{m}$ are the antennas gain and length, respectively \citep{Bale08}, $l$ is the length of antenna within the plasma cloud, and $T\,=\,2.5\,\mathrm{eV}$ is an effective temperature, of the order of the photoelectron temperature as implied by the physical process at the origin of $\delta V_\mathrm{X}$ \citep{Zaslavsky12, Pantellini12}. 
 
 On the other hand, on the Y-antenna the voltage pulse is $\delta V_\mathrm{Y}\,\approx\,\Gamma Q/C_\mathrm{sc}$, where $C_\mathrm{sc}$ is the spacecraft body capacitance which is about 200 pF \citep{Zaslavsky12}, and the charge $Q$ available in the cloud. The charge [$Q$] depends on mass, speed, angle of incidence, as well as grain and target composition. It can be determined semi-empirically with a large uncertainty with  \citep{McBride99}
 \begin{equation}\label{Q}
	Q\,\approx\,0.7m^{1.02}\,v^{3.48}
 \end{equation}
 \noindent where $Q$ is expressed in Coulombs, $m$ in kilograms, and $v$ in $\mathrm{km\,s^{-1}}$.
 
 
 However, on STEREO-B neither the X-antenna nor the Y-antenna are close enough to the preferred impact zone of fast nanometer grains, which is found to be close to the Z-antenna \citep{Zaslavsky12}. Consequently, both antennas see a voltage pulse given by $\delta V_\mathrm{Y}\,\approx\,\Gamma Q/C_\mathrm{sc}$, so that LFR can observe a dust signal only in the rare cases of impacts close to the X-or Y-antennas. Therefore on STEREO-B, LFR does not allow accurate measurements of the flux of the nanodust, but it still provides information on the rise time [$\tau$] in these rare cases.
 
 \section{Nanodust Characteristics at 1 AU}\label{s:dust}
 
 
 \subsection{Rise-Time Measurements}\label{riseData}
 
 \begin{table}
 \caption{Averaged rise time values year by year and for the whole data set for both STEREO spacecraft. $N$ is the number of spectra with detectable dust impacts, and $p$ is the percentage of the returned data that $N$ corresponds to.}\label{t:tau}
 \begin{tabular}{c c c c c c c}     
 \hline
 	 & \multicolumn{3}{c}{STEREO-A} & \multicolumn{3}{c}{STEREO-B}\\
	UC  & $N$ & $p$ [\%] & $\tau\,\pm\,\Delta\tau\ [\mu \mathrm{s}]$ & $N$ & $p$ [\%] & $\tau\,\pm\,\Delta\tau\ [\mu \mathrm{s}]$\\
 \hline
 	2007 & $158\,164$ & $20.5$ & $39\,\pm\,15$ &  $29\,008$ & $3.8$ & $40\,\pm\,22$\\
 	2008 & $287\,705$ & $36.1$ & $39\,\pm\,8$ &  $3\,030$ & $0.4$ & $43\,\pm\,22$\\
 	2009 & $124\,767$ & $15.6$ & $40\,\pm\,9$ & $36\,447$ & $4.5$ & $39\,\pm\,16$\\
 	2010 & $79\,394$ & $9.9$ & $39\,\pm\,12$ & $1\,311$ & $0.2$ & $44\,\pm\,35$\\
 	2011 & $1\,001$ & $0.1$ & $42\,\pm\,32$ & $10\,977$ & $1.4$ & $39\,\pm\,23$\\
 	all & $651\,031$ & $16.4$ & $39\,\pm\,11$ & $80\,773$ & $2.0$ & $39\,\pm\,20$\\
 \hline
 \end{tabular}
 \end{table}
 
 Our analysis of the power spectra measured by LFR allows to us determine the rise time of the pulses produced by dust-grain impacts. Since LFR does not measure individual impacts but integrates over all impacts occurring during the acquisition time, the rise-time [$\tau$] obtained in our study is an average value. 
 
 \begin{figure}
	\centerline{ \includegraphics[width=0.4999\textwidth,clip=]{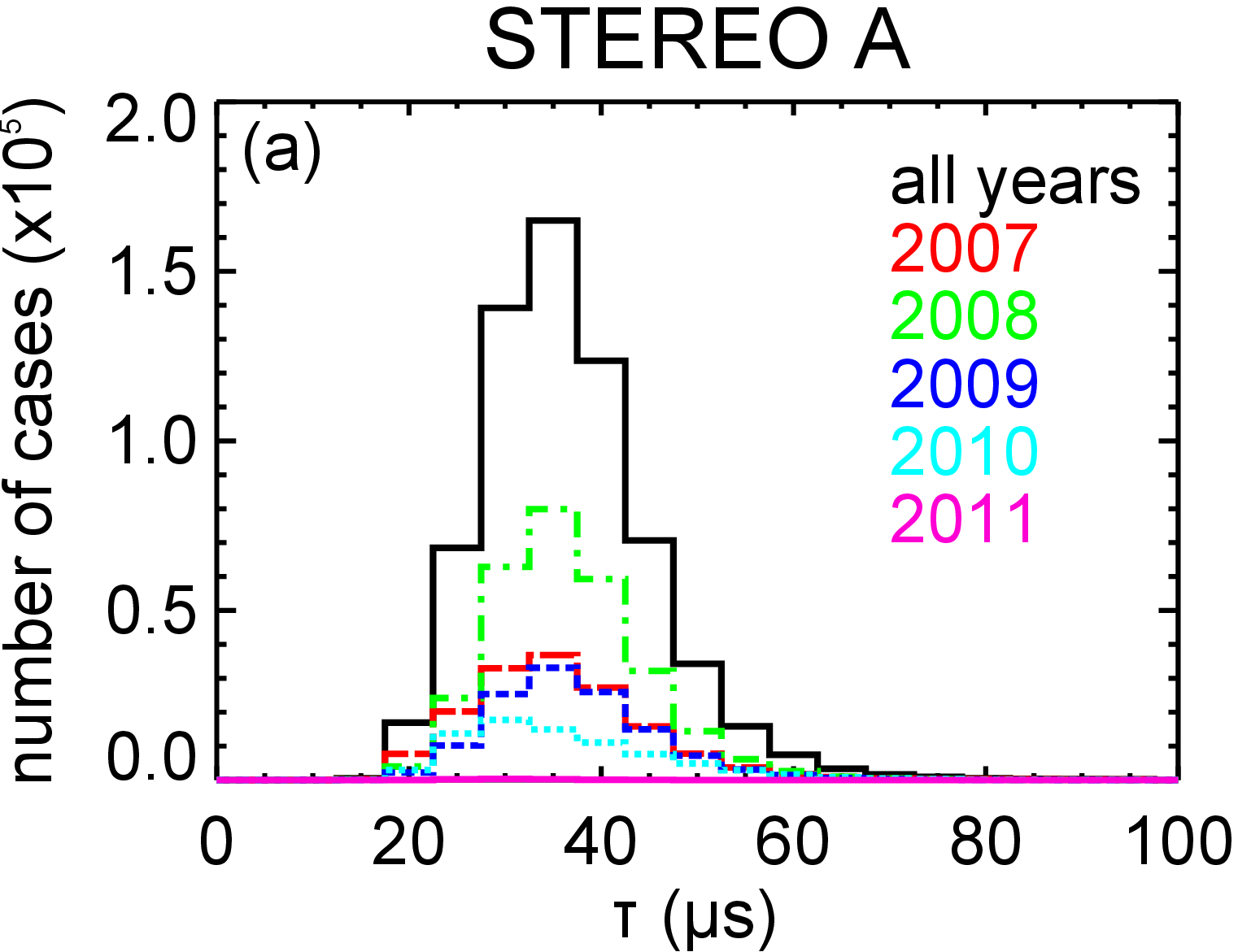} \includegraphics[width=0.4999\textwidth,clip=]{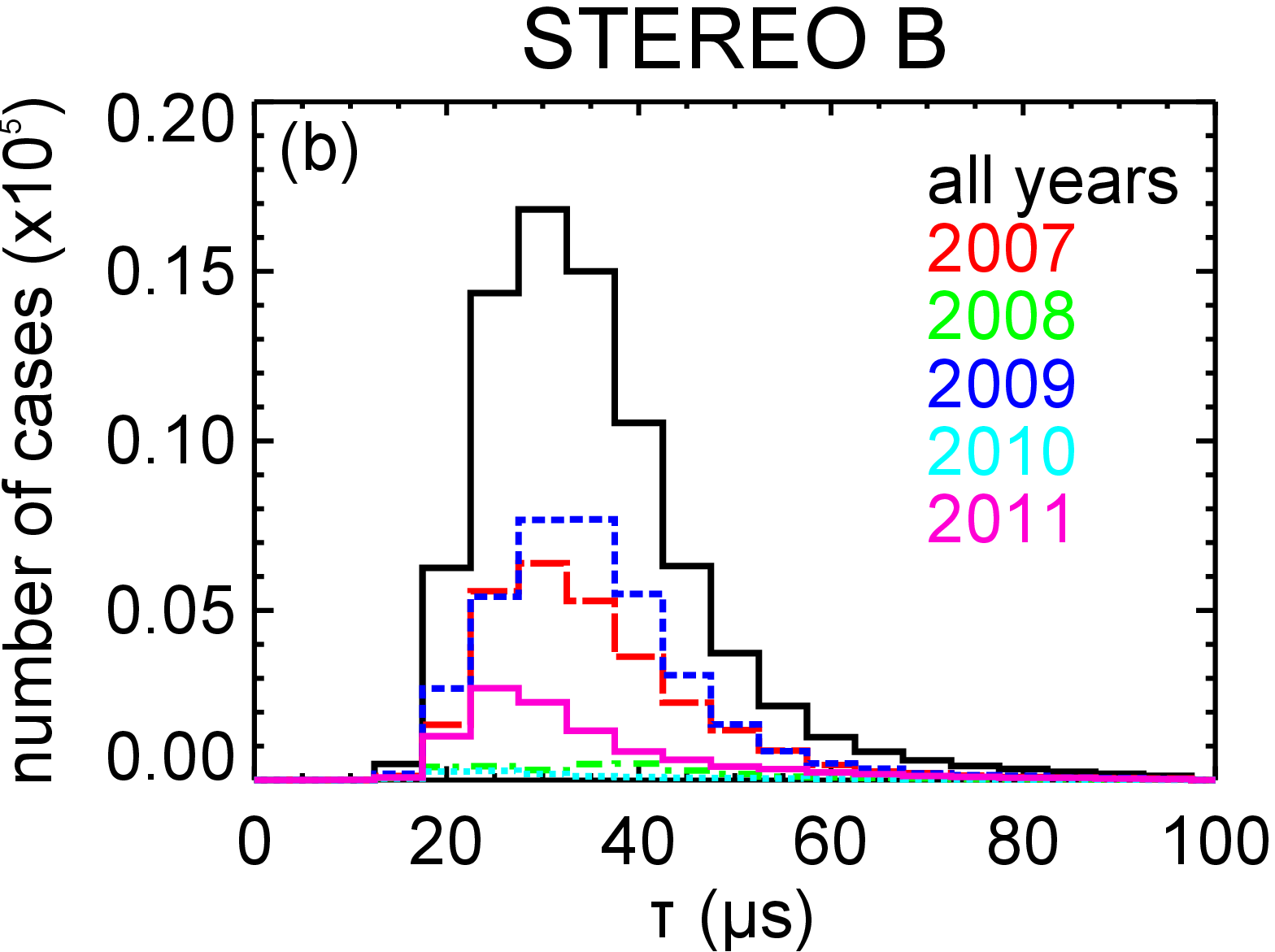}}
 	\caption{Histograms of the rise time measured by STEREO-A (panel (a)) and STEREO-B (panel (b)), for all data (black solid line) and year by year (red long-dashed line: 2007, green dotted-dashed line: 2008, blue dashed line: 2009, cyan dotted line: 2010, magenta solid line: 2011).}\label{f:tau}
 \end{figure}
 
 Table \ref{t:tau} gives the rise-time values, and Figure \ref{f:tau} presents the histograms of the rise-time measured by STEREO-A (panel (a)) and STEREO-B (panel (b)), for the whole data set and year by year. Both Figure \ref{f:tau} and Table \ref{t:tau} show that the rise time measured by LFR is stable in time and similar for both STEREO-A and B, with a value of $\tau\,=\, 39\,\pm\,12\,\mu \mathrm{s}$.  
 
The rise times measured by LFR come from the integration of all of the impacts detected during the integration time. Nevertheless, the individual measurements of the rise time made using TDS show a similar most probable value around $40\,\mu \mathrm{s}$. The difference between these two measurements is a lack of value of $\tau$ greater than $100\,\mu\mathrm{s}$ in the LFR data compared to TDS. For STEREO-A, the greatest values of $\tau$ measured by TDS are associated with the largest pulse amplitude [$V_\mathrm{X}$], corresponding to the largest dust particles. It is not surprising that the spectral analysis of LFR is not sensitive to the large dust particles, which are the less numerous ones and contribute less to the integral (Equation (\ref{v2f})) because of the large $\tau$ in the denominator.


The fact that the measured rise times are stable with time and similar for both spacecraft is in agreement with the recently proposed scenario for the interpretation of dust-impact-associated voltage pulses on boom antennas \citep{Zaslavsky12, Pantellini12b, Pantellini12}. In this scenario, the rise time is determined by the photoelectron cloud around the antenna, and can be evaluated to be proportional to the inverse plasma frequency of this photoelectron cloud. Since there are not noticeable changes in the radial distance between the Sun and the spacecraft, the plasma frequency of the photoelectron cloud is constant. Then, the value of $\tau$ is a function of the spatially varying photoelectron density  [$n_\mathrm{ph}$] inside the extended plasma cloud, which is difficult to quantify \citep{Pantellini12b}. From our observed value of $\tau$, the ``mean'' value of $n_\mathrm{ph}$ in the cloud is about $10\,\mathrm{cm}^{-3}$. This scenario also implies that the expansion time of the plasma cloud from the impact zone to its maximum size (\textit{i.e.} when the plasma cloud density reaches the value of the ambient solar-wind density) is greater than the measured value of $\tau$.

 \subsection{Determination of the Flux from LFR Measurements}\label{s:flux}
 
 
 The parameters [$\tau$ and $\langle N\delta V^2 \rangle$] obtained from the fitting of the LFR spectra in Band A of STEREO-A enable us to infer the nanodust flux observed at 1 AU. 
Considering a dust grain impacting at a distance $r$ from the antenna, and a spherical expansion of the impact cloud, the length of the antenna within the cloud [$l$] is given by $l\ =\ \sqrt{R_\mathrm{C}^2\,-\,r^2}$ where $R_\mathrm{C}\,\approx\,(3Q/4\pi e n_\mathrm{a})^{1/3}$ is the maximum size of the plasma cloud. The maximum value of the distance $r$ for each dust impact is the minimum value between $R_\mathrm{C}$ and $R_\mathrm{SC}$, the order of magnitude of the spacecraft size \citep{Zaslavsky12}. Then assuming a cylindrical symmetry around the antenna, so that the probability that an impact occurs at a distance between $r$ and $r\,+\,\mathrm{d}r$ from the antenna is proportional to $2\pi r \mathrm{d}r$, we can write
 \begin{equation}\label{ndvFLux}
\langle N \delta V^2/\tau^2\rangle \ \approx\ \int_{m_\mathrm{min}}^{m_\mathrm{max}}f(m) \,\int_{0}^{\mathrm{min}(R_\mathrm{c},R_\mathrm{SC})} \frac{\delta V^2}{\tau^2} 2\pi r \mathrm{d}r \mathrm{d}m
 \end{equation}
 
 \noindent where $f(m)\mathrm{d}m\,=\,f_0m^{-\gamma}\mathrm{d}m$ is the flux of particles of mass between $m$ and $m\,+\,\mathrm{d}m$. Note that the interplanetary dust models would give $f_0\,=\,7.4\,\times\,10^{-19}\,\mathrm{kg^{-1}\,m^{-2}\,s^{-1}}$, and $\gamma\,\approx\,11/6$ in the mass range $\left[m_\mathrm{min}\,:\,m_\mathrm{max} \right]$ \citep{Grun85, Ceplecha98}.
 
  To obtain the nanodust flux from Equation (\ref{ndvFLux}), we need the average value of the squared impulse voltage produced by a cloud of charge $Q$: $\delta V^2$. Using Equation (\ref{deltaV}) for $\delta V$ and considering both cases of a plasma cloud larger and smaller than the spacecraft effective size
\begin{equation}\label{intV2}
	\langle \delta V^2 \rangle  = \left\{ 
  \begin{array}{l l}
		\frac{\Gamma^2 T^2}{L^2} \frac{R_\mathrm{C}^4}{2 R_\mathrm{SC}^2} & \quad \mathrm{if}\ R_\mathrm{C} < R_\mathrm{SC} \\
		\frac{\Gamma^2 T^2}{L^2} \left( R_\mathrm{C}^2\,-\,\frac{R_\mathrm{SC}^{2}}{2} \right) & \quad \mathrm{if}\ R_\mathrm{C} > R_\mathrm{SC} \\
	\end{array} \right. .
\end{equation}
\noindent In the mass range considered in this study and using Equation (\ref{Q}), we have $R_\mathrm{C}\,\approx\,Km^{1/3}$, with $K\,\approx\,(3\times0.7 v^{3.5}/4\pi n_\mathrm{a} e)^{1/3}\,\approx\,5\times 10^6\,\mathrm{m\,kg}^{-1/3}$, for typical values of the ambient solar-wind density $n_\mathrm{a}$, and impact speed $v\,=\,300\,\mathrm{km\,s^{-1}}$ \citep{Mann2010}, which is close to the solar wind velocity because of the pick-up-ion-like mechanism. For STEREO-A,  the effective sensitive area is $R_\mathrm{SC}^2\,\approx\,0.7\,\mathrm{m}^{2}$ \citep{Zaslavsky12}. The mass for which $R_\mathrm{SC}\,=\,R_\mathrm{C}$ is $m_\mathrm{R_{SC}}\,=\, 1.6\times 10^{-20}\,\mathrm{kg}$. This mass is of the same order of magnitude of the maximum dust mass detected, $m_\mathrm{max}\,\approx\,2\times10^{-20}$kg, corresponding to the mass range measured by TDS \citep{Zaslavsky12}.



Assuming that $\tau$ is a constant, according to Section \ref{riseData}, we integrate Equation (\ref{ndvFLux}) using Equation (\ref{intV2})

\begin{eqnarray}\label{fluxNVT_step1}
	\langle N \delta V^2/\tau^2\rangle & \simeq & f_0 \pi R_\mathrm{SC}^2 \left( \frac{\Gamma T}{L\tau}\right)^2 \left( \left[\frac{K^4 m^{-\gamma+7/3}}{2R_\mathrm{SC}^2(-\gamma + 7/3)} \right]_{m_\mathrm{min}}^{m_\mathrm{R_{SC}}} \right.\nonumber \\
	 & & + \left. \left[\frac{K^2m^{-\gamma + 5/3}}{-\gamma + 5/3}\, -\,\frac{R_\mathrm{SC}^2 m^{-\gamma + 1}}{2(-\gamma + 1)} \right]_{m_\mathrm{R_{SC}}}^{m_\mathrm{max}} \right)
\end{eqnarray}

Assuming that $m_\mathrm{min} \ll m_\mathrm{R_{SC}} = m_\mathrm{max}$, and taking $\gamma = 11/6$, Equation (\ref{fluxNVT_step1}) becomes

\begin{equation}\label{fluxNVT}
	\langle N \delta V^2/\tau^2\rangle\ \approx\ f_0 \pi K^4 m_\mathrm{R_{SC}}^{1/2} \left( \frac{\Gamma T}{L\tau}\right)^2,
\end{equation}

\noindent which only depends on the mass $m_\mathrm{R_{SC}}$. 

\subsection{Nanodust Fluxes at 1 AU}

 \begin{figure}
	\centerline{ \includegraphics[width=0.9\textwidth]{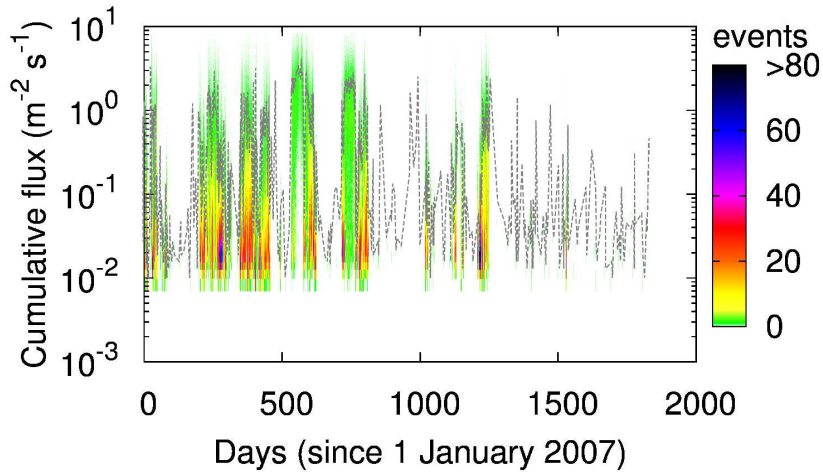}}
 	\caption{Variation with time of the cumulative flux of particules of mass greater than $10^{-20}$kg measured by STEREO-A/WAVES LFR between 2007 and 2011. The color scale of the cumulative flux distribution has been chosen to emphasize the daily most-probable value of the flux. The dotted gray line corresponds to the daily mean of the measured nanodust flux.}\label{f:fluxTime}
 \end{figure}

Figure \ref{f:fluxTime} represents the cumulative flux of particles of mass greater than $10^{-20}$kg ($F_{10^{-20}}$), measured by STEREO-A between 2007 and 2011 using Equation (\ref{fluxNVT}). The cumulative flux is used to allow direct comparison with previous studies and interplanetary-dust models, and is given by
\[ F_\mathrm{m}\ =\ \int_m^{+\infty} f(m)\mathrm{d}m\ =\ \frac{f_0\,m^{1-\gamma}}{\gamma-1}
\]

\noindent The flux has large fluctuations, as first noted by \citet{Meyer09} and confirmed by \citet{Zaslavsky12}. The white periods on Figure \ref{f:fluxTime} correspond to time periods when STEREO-A/WAVES LFR did not measure dust.  This is expected to be due, at least in part, to geometrical effects. One clue for the geometric origin is inferred by the quasi-continuous flux measured by TDS on STEREO-B \citep{Zaslavsky12}. Another indication of the geometrical origin of these fluctuations is the dependence of the flux measured by STEREO-A with the latitude in ecliptic coordinates.  It is noteworthy than this behavior is not seen by STEREO-B (D. Malaspina, private communication, 2012). 





\begin{figure}
     \centerline{ \includegraphics[width=0.8\textwidth,clip=]{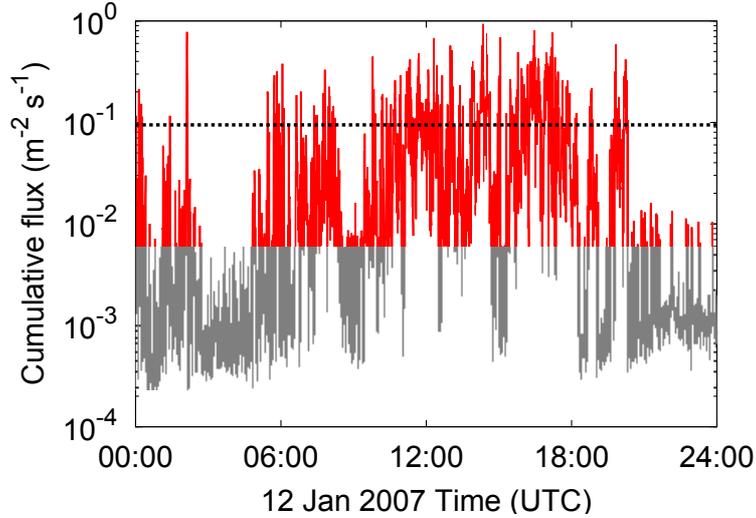}}
     \caption{Variation during 12 January 2007 of the STEREO-A/WAVES LFR measurements. The data for which the thermal noise is dominant, so that the dust flux cannot be measured, are plotted in gray, whereas the genuine dust flux measurements are plotted in red. The black dotted line shows the average cumulative flux [$F_{10^{-20}}$] of nanodusts measured on 12 January 2007.}\label{f:daily}
\end{figure}


In addition to the temporal variations of the nanodust flux at large time scales, the high temporal resolution of LFR enables us to study the fast variations of the nanodust flux detected by STEREO-A. Figure \ref{f:daily} shows such variations of the cumulative flux, $F_{10^{-20}}$, during 12 January 2007. In this example, $46\,\%$ of the spectra measured by LFR show nanodust impacts, with long continuous periods of either dust or thermal-noise measurements, and periods of fast alternation between both (as between 19:00 and 21:00). These three behaviors take place at different time scales, from a few minutes up to days, and the longest continuous measurement of dust in our data set is 11.5 days starting on  13 January 2008.

\begin{figure}
	\centerline{ \includegraphics[width=0.8\textwidth,clip=]{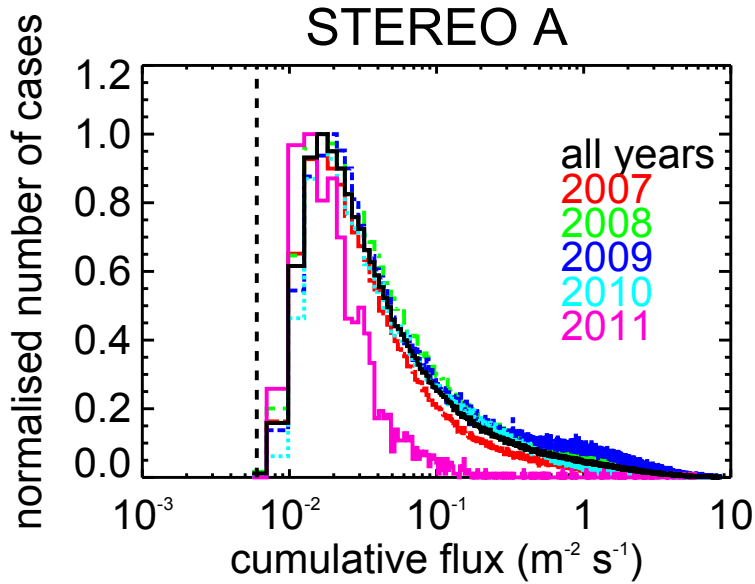}}
 	\caption{Histograms of the cumulative flux of particles of mass greater than $10^{-20}$kg measured by STEREO-A or all data (black solid line) and year by year (red long-dashed line: 2007, green dotted-dashed line: 2008, blue dashed line: 2009, cyan dotted line: 2010, magenta solid line: 2011), normalised to their respective maximum values. The dashed line represents the instrumental limits of dust detection. Note the log-scale on the $x$-axis.}\label{f:flux}
 \end{figure}

Figure \ref{f:flux} shows the histograms of $F_{10^{-20}}$ measured by STEREO-A, when LFR measures dusts. The histograms are normalized to their respective maximum. One can see in Figure \ref{f:flux} that the distribution of measurement is stable at large temporal scales, even if the number of measurements can vary strongly from one year to another (from $287\,706$ in 2008 to only $1\,001$ in 2011, see Table \ref{t:tau}). These measurements agree with the TDS measurement \citep{Zaslavsky12}, with the previous study of LFR measurements \citep{Meyer09}, and interplanetary dust distribution models, in which $F_{10^{-20}}\,=\,0.04\ \mathrm{m^{-2}\,s^{-1}}$ \citep{Grun85, Ceplecha98}. The most-probable value of the flux is not much larger than the instrumental limit of detection, leading to a very asymmetric distribution. Even if LFR could measure smaller flux, it is noteworthy that flux lower than $2\times 10^{-3}\ \mathrm{m^{-2}\,s^{-1}}$ cannot be measured due to the plasma thermal noise, as shown in Figure \ref{f:daily}. Since the geometrical effects, like projection effects including the flow direction of the dust and the orientation of the spacecraft, on dust detection are not fully understood at this time, it is not possible to determine the origin of the large difference of dust measurements from one year to another.




 \section{Conclusions}

Even if not designed to do so, S/WAVES LFR allows accurate measurements of the interplanetary nanodust flux at a high temporal resolution. Nevertheless, the detection mechanism needs one of the antenna booms used by the receiver to be inside the plasma clouds created by the dust impacts. Consequently, only STEREO-A can be used to determine the nanodust flux with LFR. However, complementary measurements by TDS allow one to study dust with STEREO-B. The complementarity of the time- and frequency-domain measurement of nanodust using antenna booms is not limited to this point.  Both TDS and LFR instruments measure nanodust signals with a similar rise time of $40\,\mu$s and a similar flux, but with complementary advantages in dust flux measurements. TDS measures individual nanodust impacts in monopole mode using the three antennas, allowing dust measurement on both STEREO spacecraft but with a limited telemetry, whereas LFR gives a continuous survey at a high time resolution. 

Geometrical effects on nanodust detection using S/WAVES have been pointed out in this study and previous ones, but due to the very complex geometry of the spacecraft from the point of view of the antenna booms and in the moving frame of the dust, we do not yet fully understand the observed variations of the nanodust flux.  Nevertheless, this continuous survey of the nanodust in the interplanetary medium at high time resolution allows the ongoing study of the correlation between the local solar-wind plasma and the properties of the dust. In addition to the TDS measurements, this study should lead to important insights on dust detection mechanisms for both STEREO and nanodust acceleration by the magnetized solar wind.

The spectral analysis of more than $700\,000$ samples shows that the rise time of the signal created by dust impacts on the spacecraft is stable, in agreement with the recently proposed scenario for the interpretation of the dust impact associated voltage pulses on boom antennas \citep{Zaslavsky12, Pantellini12b, Pantellini12}. This enables us to define the spectral and temporal resolution and domain required to measure accurately dust with an antenna boom for future instruments, such as \textit{Fields} on \textit{Solar Probe Plus}. 

From the insight on the detection mechanism, we have updated the previous determination of the dust flux measured by LFR \citep{Meyer09}. With our improved interpretation, we found a flux of the same order of magnitude, but without requiring estimates of unknown parameters, such as the expansion velocity of the cloud or the minimal mass of impacting dust. The main uncertainty of our measurements comes from  the relation between the released charge, the mass, and the impact speed (Equation (\ref{Q})) for nanodust. It is noteworthy that this limitation is inherent to all measurements of dust in this size range, and that ongoing and future improvements of dust accelerators in laboratory experiments should resolve this issue in the future. In this study, we have assumed that the power-law index of the interplanetary dust distribution models \citep{Grun85, Ceplecha98} is valid in the nanometer range, in other words that the nanodust flux behaves as if these particles were created by collision equilibrium. Further analysis of the TDS measurements of nanodust could infirm this, leading to an observed distribution of the nanodust flux. In such a case, the numerical values of the flux that we give in the present study will have to be changed by updating the value of $\gamma$ in Equation (\ref{fluxNVT_step1}).


%

%

%

%
 \begin{acks}
We thank the team who designed and built the instrument. The French part was supported by CNES and CNRS. 
 \end{acks}

%
%
 \bibliographystyle{spr-mp-sola}
 \bibliography{nanoDustLFR}  
%
%
%
%

\end{article} 
\end{document}